

\magnification=\magstep1

\centerline{\bf Errata in Binney and Tremaine, ``Galactic Dynamics''}

\medskip

\centerline{March 1993}

\bigskip

\noindent
This list does not include minor or obvious typographical errors, except that
all known typos in mathematical formulae are included, no matter how small.
The errors are divided into two classes, ``Potentially dangerous errors''
(mathematically incorrect statements or seriously misleading errors in
equations), and ``Innocuous errors''.

The TeX file of this list is available from the authors at the e-mail
addresses below and will be updated if more errors are discovered.

Some of these errors may be corrected in later printings of the book.

We are grateful to the following colleagues to pointing out many of these
errors: David Earn, Stefan Engstr\"om, Gerry Gilmore, Chris Hunter, Doug
Johnstone, Blane Little, Thomas Lydon, Phil Mahoney, Kap-Soo Oh, Sterl
Phinney, Noam Soker, S. Sridhar, Rosemary Wyse, and Harold Zapolsky.
Additional contributions may be sent to the authors at the e-mail addresses:

\noindent
{\tt binney@zeus.thphys.ox.ac.uk}

\noindent
{\tt tremaine@cita.utoronto.ca}

\bigskip

\noindent
{\bf Potentially dangerous errors}

\smallskip

\itemitem{p. 45} Contour diagrams in Figure 2-6 are incorrect.

\itemitem{p. 57} Expression for $I$ for prolate spheroid should read
$${1\over e}\ln\left(1+e\over 1-e\right).$$
In the original there is an incorrect extra factor of $\sqrt{1-e^2}$.

\itemitem{p. 102} In the first line of eq.\ (2P-15), the factor $2\pi$
preceding the summation symbol should be replaced by $2G$. Also,
eq.\ (2P-16) should read
$$\alpha_k=\pi\left[(2k)!\over 2^{2k}(k!)^2\right]^2;$$
the original erroneously used $2^{k}$ instead of $2^{2k}$.

\itemitem{p. 110} The first line of eq.\ (3-35) should read
$$\Delta\psi=2L\int_{s_1}^{s_2}{dI\over r^2}={2L\over
b\sqrt{-2E}}\int_{s_1}^{s_2}{(s-1)ds\over s(s-2)\sqrt{(s_2-s)(s-s_1)}}.$$

\itemitem{p. 140} The sentence following eq.\ (3-106) is incorrect. It is
true that $\Phi_{xx}+\Phi_{yy}+4\Omega_b^2$ is positive, but it does not
follow that $L_4$ and $L_5$ are always stable. To correct the error, add a
period at the end of eq. (3-106) and replace the rest of the paragraph
by: Deciding whether condition (3-100b) holds is tedious in the general case,
but straightforward in the limit of negligible core radius,
$(\Omega_bR_c/v_0)\to0$ (which applies, for example, to Figure 3-13). Then it
is easy to show that (3-100b) holds---and thus that $L_4$ and $L_5$ are
stable---providing $q^2>5[({32\over25})^{1/2}-1]\simeq(0.81)^2$.

\itemitem{p. 550} Eq.\ (8P-9) has three misprints: the plus sign in front of
$A$ should be a minus; the denominator under $A$ should be 8, not 4, and the
exponent of $\widetilde r$ beside $A$ should be $-{1\over 2}$, not ${1\over
2}$. Thus the corrected equation reads
$$x\equiv {r_b\over GM\beta}\simeq{\textstyle {1\over2}}
-{A{\widetilde r}^{-{1\over2}}\over8}
\left[\cos\big({\textstyle {1\over2}}
\surd{7}\ln\widetilde r+\phi\big)+\surd{7}\sin\big({\textstyle {1\over2}}
\surd{7}\ln\widetilde r+\phi\big)\right]$$.

\itemitem{p. 658} In eq.\ (1C-51), the upper limit of the integral should be
$\pi$, not $\infty$.

\bigskip

\noindent
{\bf Innocuous errors}

\smallskip

\itemitem{p. 31} First term on right-hand side of eq.\ (2-7) is missing a
superscript and should read $-3/|{\bf x}'-{\bf x}|^3$.

\itemitem{p. 36} First line after eq.\ (2-22) should read ``From Newton's
{\it first and} second theorems or from equation (2-22) it follows$\ldots$''.

\itemitem{p. 38} Factor $G$ in denominator of eq.\ (2-35) should be
deleted.

\itemitem{p. 44} Sentence three lines above eq.\ (2-51) should begin ``For
example, if we take the $(n-1)$st derivative of $\Phi_K(R,z)/a$ with respect to
$a^2\ldots$''.

\itemitem{p. 55} Second line following eq.\ (2-81) should finish
``$\ldots$implies $\sinh u_m=\sqrt{(1-e^2)}/e$.'' In the original the factor
$e$ is squared, which is incorrect.

\itemitem{p. 63} Factor $1/R$ in front of first term in eq.\ (2-108a)
should be deleted.

\itemitem{p. 66} In the first line of eq.\ (2-122), the lower limit on the
sum should be $a=r$, not $r=a$.

\itemitem{p. 75} Second last line before eq.\ (2-155) should read ``the
surface density $\Sigma_k(R)$ of the sheet$\ldots$''.

\itemitem{p. 78} Last word on the page should be ``dotted'', not
``dashed''.

\itemitem{p. 100} Right-hand side of the formula at the end of the second
line of the page should read $a^2(\cosh u+|\cos v|)^2$. In the original the
factor $a$ was not squared.

\itemitem{p. 101} Lower limit of the integral in eq.\ (2P-9) should be
$R$, not $r$.

\itemitem{p. 108} Sentence just after eq.\ (3-23) should read
``$\ldots$since $r\to\infty$ as $(\psi-\psi_0)\to\hbox{arccos}(-1/e)$; the
orbit$\ldots$''; in the original the minus sign is missing.

\itemitem{p. 109} Last equality in eq.\ (3-33) should read
$$={2\pi b\over\sqrt{-2E}}\big[{\textstyle {1\over2}}
(s_1+s_2)-1\big].$$

\itemitem{p. 138} In the second part of eq.\ (3-94) there should be a dot
over $\xi$, that is, $\ddot\eta=-2\Omega_b\dot\xi-\Phi_{yy}\eta$.

\itemitem{p. 139} Last term on the left side of eq.\ (3-101) should be
$16\Omega_b^4$, not $16\Omega_b^2$.

\itemitem{p. 147} Factor $R$ under square root in eq.\ (3-111) should be
replaced by $1/R$.

\itemitem{p. 173} Reference in the caption of Figure 3-27 should be to
eq.\ (3-159), not (3-161).

\itemitem{p. 184} The symbol $\Omega$ in eq.\ (3P-2) should be replaced by
$\Omega^2$.

\itemitem{p. 185} In eq.\ (3P-5) there should be a minus sign before the
arctan.

\itemitem{p. 196} First line after eq.\ (4-22) should read ``The last term
on the {\it left} side can be$\ldots$''. Also, in the last line before
eq.\ (4-25), $v_j$ should be replaced by ${\overline v}_j$.

\itemitem{p. 198} Last sentence preceding eq.\ (4-31) should read
``$\ldots$we may evaluate equation (4-29a) at $z=0$,$\ldots$''. In the
original the equation number is missing.

\itemitem{p. 214} In eq.\ (4-83), $v_x^2$ should be $\overline{v_x^2}$.

\itemitem{p. 220} In eq.\ (4-100), the factor following the summation
should be $(\partial f/\partial I_m)(d I_m/dt)$; in the original the subscript
was incorrectly given as $n$.

\itemitem{p. 225} In eq.\ (4-113), $\psi$ should be replaced by $\Psi$;
that is, the first equality should read $\rho=c_5\Psi^5$.

\itemitem{p. 272} Twelfth line in third paragraph should read ``pump energy
from the most {\it massive} particles$\ldots$''.

\itemitem{p. 300} In the second line following eq.\ (5-48), the phrase
``in square brackets'' should be deleted.

\itemitem{p. 381} In last line of the caption to Figure 6-20, replace
``merging'' by ``emerging''.

\itemitem{p. 424} Eight and ninth lines should read ``by Newton's first {\it
and second} theorems$\ldots$''

\itemitem{p. 435} First line following eq.\ (7-41) should read ``In any
{\it static} axisymmetric system$\ldots$''.

\itemitem{p. 469} Last line, delete the word ``spherical''.

\itemitem{p. 487} In eq.\ (7P-11), the factor $\omega$ should be inside the
integral sign.

\itemitem{p. 488} In eq.\ (7P-14) the right bracket after the
term $A/\Omega$ should be removed.

\itemitem{p. 502} In the first line following eq.\ (8-42) replace ``second
line'' by ``second equality''. In the first line after eq.\ (8-43) replace
``equation (4-123a)'' by ``equation (4-125a)''.

\itemitem{p. 534} In eq.\ (8-106), the first factor inside the square brackets
should be $(m_1+m_a)$, not $(m+m_a)$.

\itemitem{p. 643} The units of the solar mass are g, not cm.

\itemitem{p. 677} In eq.\ (4A-2a), $\overline{f}({\bf x},{\bf v},t)$ should be
replaced by $\overline{f}({\bf x},{\bf v})$.

\itemitem{p. 690} In eq.\ (6A-4) the argument of $\Phi_1$ should be $({\bf
x}',{\bf v}',t')$, that is, a prime should be added to $t$.

\bye